\documentclass{iau}
\usepackage{graphicx}
\usepackage{bm}
\usepackage{empheq}

\title[Chromospheric diagnosis with forward scattering polarization] 
{Chromospheric diagnosis with forward scattering polarization}

\author[E. S. Carlin]   
{E. S. Carlin
}

\affiliation{Istituto Ricerche Solari Locarno, Monti Locarno 6600, 
Switzerland\\ email: {\tt edgar@irsol.ch}} 

\pubyear{2015}
\volume{305}  
\jname{Polarimetry: From the Sun to Stars and Stellar Environments}
\editors{K.N. Nagendra, S. Bagnulo, \\ R. Centeno, \& M.J. Mart\'inez Gonz\'alez, eds.}
\begin{document}

\maketitle

\begin{abstract}
Is it physically feasible to perform the chromospheric diagnosis using spatial maps of scattering polarization at the solar disk center? To investigate it we synthesized polarization maps (in $8542$ {\AA}) resulting from MHD solar models and NLTE radiative transfer calculations that consider Hanle effect and vertical macroscopic motions. After explaining the physical context of forward scattering and presenting our results, we arrive at the definition of Hanle polarity inversion lines. We show how such features can give support for a clearer chromospheric diagnosis in which the magnetic and dynamic effects in the scattering polarization could be disentangled.
 
\end{abstract}

\firstsection 

\section{Introduction}
\label{sec-intro}

The diagnosis of solar magnetic fields is possible by modeling and understanding the spectral line polarization signals created by scattering processes that result from the matter-radiation interaction. We aim at improving the diagnosis strategies for the chromosphere (namely, for the Ca\,{\sc ii} $8542$ {\AA} line) by investigating the spatial behavior of such scattering polarization when it emerges from the solar disk center, instead of following the usual approach of studying the solar limb \cite[(see also {{Bianda} {et~al.} 2011}; Anusha et~al. 2011)]{Anusha11,michele2011cai}. To understand the differences between both extreme cases we start summarizing the factors affecting the spectral line polarization in a stellar atmosphere. 

A first source of polarization is the local magnetic field $\mathbf{B}$ permeating the medium. It interacts locally with the atomic system during the scattering process and can thus modify the polarization through the Zeeman and Hanle effects \cite[(Bommier\ 1997; Landi Degl'Innocenti \& Landolfi\ 2004, LL04 hereafter)]{{bommier97},ll04}. In particular, the relatively weak fields of the quiet chromosphere make the linear polarization (LP) signals of the $8542$ {\AA} line to be controlled by the Hanle effect while its circular polarization is produced by the longitudinal Zeeman effect. In the disk center (forward scattering geometry), the Hanle effect creates LP due to the symmetry breaking produced by inclined magnetic fields \cite[({{Trujillo Bueno} 2003})]{Trujillo-Bueno:2003aa}.

A second factor altering the polarization is the anisotropy of the radiation field ($\mathcal{A}$) illuminating the emitting plasma differently in each spatial point along the line of sight \cite[({Trujillo Bueno} 2001; {Holzreuter} {et~al.} 2005)]{{jtb01},{Holzreuter:2005aa}}. The radiation field anisotropy can change strongly in space and time because it is mainly modulated by vertical gradients of temperature and velocity (dynamics hereafter). Indeed, recent investigations come to the conclusion that, due to the strong chromospheric kinematics, the anisotropy has the largest potential for modifying the LP amplitudes \cite[({{Carlin} {et~al.} 2013})]{Carlin:2013aa}. Neglecting it could make us attribute wrongly polarization amplitudes and profiles to quantum effects, thus leading to erroneous diagnosis of the second solar spectrum \cite[(SSS, see Stenflo \& Keller 1997)]{sk97}. 

Collisions are a third factor to meet \cite[(e.g., Bommier 2009)]{Bommier:2009aa}. They are essentially isotropic and hence depolarizing, with more efficiency deeper in the atmosphere due to the larger density of colliding particles. 

In last place, the curvature of the solar surface can also be considered as a polarization driver because it naturally modifies the observer's perspective: the farther an observed point is from the disk center (larger heliocentric angle) the larger the inclination of the LOS ($\theta$) with respect to its local solar vertical. This generates the following two effects altering the diagnosis.

 The first effect is a different response of the polarization to the solar conditions according to the LOS. This can be understood considering a generic LOS with $\mu=\cos(\theta)$ and expressing the Stokes Q line-center emissivity (LL04, pag. 290) as:
\begin{equation}\label{eq:one}
\centering 
\epsilon^l_{Q} (\nu_0, \mu) \propto (1-\mu^2)f_1(\mathcal{A},\mathbf{B},C) + \mu \sqrt{1-\mu^2}f_2(\mathcal{A},\mathbf{B},C)+(1+\mu^2)f_3(\mathcal{A},\mathbf{B},C).
 \end{equation}
 The functional form of this equation results from quantum electrodynamics laws\footnote{Note that the polarization scattered by a plasma element is intrinsically dependent on the LOS, no matter whether the plasma is in the Sun or in a laboratory. Equation (\ref{eq:one}) simplifies the mathematical formulation while highlighting the polarization drivers $(\mathcal{A},\mathbf{B}, C)$ on which the $f_i$ functions generally depend in the solar case. Actually, such functions depend first on the atomic density matrix, which is tight to solar conditions through the statistical equillibrium equations.}. Applied to the Sun, each $\mu$ factor weights different addends indirectly depending on the local solar conditions (hidden in the functions $f_i$ for simplicity). The LOS thus sets the way the LP reacts to other polarization drivers, such as the magnetic field, by weakening or strengthening each contribution. The opposite is also true: the solar conditions sizing the $f$ functions set out the governing $\mu$ factor. Namely, the strongest contribution to the polarization is usually given by $f_1$ because it is indirectly enhanced by vertical symmetry breakings in the radiation field such as the limb darkening (contrarily to $f_2$ and $f_3$, which depend on weaker terms generated by azimuthal symmetry breakings around the solar vertical as the ones produced by inclined magnetic fields). As a result, the essential dependence of the scattering polarization amplitudes on the LOS is roughly described by the factor $(1-\mu^2)$: the LP increases towards the solar limb. Exceptions to this rule can come from relative increments in the other addends of Eq. \ref{eq:one} (e.g., due to particular magnetic configurations). Our goal here will be to understand what happens when the dominating factor naturally vanishes, i.e. in $\mu=1$, where the solar curvature does not impose a preferential direction for the polarization.  

\cite[{There are other issues related to the solar curvature (e.g., {Mili{\'c}} \& {Faurobert}, 2012})]{Milic:2012aa} but here we concentrate on a second effect in relation to the LOS, which is the change of the region where the spectral line forms. Comparing with an observation closer to disk center, the light emerging from the limb follows a geometrical path inside the solar atmosphere that is longer, higher, and crossing a shorter height range before reaching optical depth unity\footnote{A generic LOS ($\mu$) entering in a plane-parallel atmosphere at a given height H arrives at $\tau_{\nu_0}\approx 1$ at a height $h=H-\ln(\mu+1)$, after a geometrical path $\Delta s=\ln(\mu+1)/\mu$ (distances in scale height units).}. 
For a chromospheric line, a higher formation layer means lower depolarizing collisional rates with neutral atoms as well as larger temperatures (hence larger emissivities and ion densities) and larger velocity/temperature gradients (hence larger radiation field anisotropy). Thus, a  LOS inciding more parallel to the solar surface collects more polarized photons because it crosses a larger superposition of scatterers that are furthermore in optimal physical conditions for creating polarization.

Both LOS effects increase the scattering polarization toward the limb, sourcing the SSS. But the problem with close-to-limb diagnosis is that the LOS crosses many solar radii that are horizontally uncorrelated, so that their physical state can be considered as varying randomly along the optical path. In the case of local MHD models, the {horizontal} variations also depends on the size of the simulation domain insofar as periodic sick-boundary conditions are usually applied. Statistically speaking, it is more difficult for a model atmosphere to reproduce the horizontal spacial variations of the Sun than the vertical ones. As solar and synthetic spatial variations are encoded in wavelength in the radiative transfer, a statistical reconciliation between both sources is required. 
If not, the degeneracy in explaining the spectral line profiles measured/synthesized near the limb is increased too much due to the large horizontal variability of the solar atmosphere, the longer and horizontally extended formation region, and, in the current solar models, the imperfect statistical behavior of the models. On the contrary, hydrostatic equilibrium and vertical stratification are dominant ingredients that seem to assure a sufficient degree of physical correlation and realism along the vertical. Hence, the corresponding variations are essentially predefined by the density stratification and the kinematic boundary conditions, which can be emulated from photospheric dopplergrams as done in \cite[{{Carlsson} \& {Stein} (1997)}]{carlsson_stein97}. This physics is contained with a presumably good level of realism in the solar models but to capture such advantage radiatively, without the undesired effects of mixing information from different solar verticals, a radial LOS is required.

The previous ideas suggest that the disk center geometry could be more suitable for diagnosing the chromosphere.
Briefly, its advantages are: (i) although the LP is weaker (due to the cancellation of the dominant contribution in Eq 1.1), the number of collected photons is significantly increased with respect to the limb (limb darkening); (ii) the geometry maps the Stokes vector of single pixels to a single vertical stratification in the atmosphere, which minimizes RT effects expected from horizontal inhomogeneties and avoids mixing of horizontal structures along the LOS; (iii) in contrast to slit-like observations, the spatial continuity in a disk center map simplifies the interpretation of the magnetic topology and, as made clear in next sections, discriminates the magnetic contributions to the LP signals from the kinematic ones; and (iv), partial redistribution (PRD) effects in the scattering are minimized at disk center, which supresses polarized emission in the spectral wings \cite[({{Stenflo} 2006})]{Stenflo:2006aa}. 

Thus, avoiding PRD, the mixing of horizontal structures along the LOS and the polarization introduced by the solar curvature, the LP profiles at disk center are ``purer''. They are essentially driven by magnetic field (Hanle effect in quiet Sun) and solar dynamics (via radiation field anisotropy). The fundamental point is the discrimination of these two drivers. This is possible by ``realistically'' simulating and understanding the variations of the LP signals due to dynamic effects.
\section{Radiative transfer calculations}
\label{sec1}
The polarization maps shown in the next section were obtained solving the RT problem in a radiation MHD simulation of the solar atmosphere computed by \cite[Leeanaarts et al. (2009)]{Leenaarts:2009} with the Oslo Stagger Code \cite[(Hansteen et al. 2007)]{Hansteen:2007}. The snapshot represents the quiet Sun ($\left\langle \mathrm{B (300\,km}) \right\rangle = 120$ G), with a size $\Delta x \times \Delta y \times \Delta z = 5.85 \times 5.98 \times 4$ Mm.

 We initialized the RT using the corresponding 3D NLTE Ca {\sc ii} level populations provided by \cite[Leenaarts et al. (2009)]{Leenaarts:2009}. During the RT each model column was treated as an independent plane-parallel atmosphere (hence neglecting the effect of horizontal inhomogeneities in the plasma) and the solution for the Stokes vector was obtained self-consistently (iteratively) with the solution to the statistical equillibrium equations (SEE) for the multipolar tensor components of the atomic density matrix $\rho^K_Q(\mathrm{J})$ (with $K=0,...,2\mathrm{J}$ and $-K\leq Q \leq K$  in each energy level J). Expressions of the RT coefficients and radiation field tensors are similar to the ones in \cite[Manso Sainz \& Trujillo Bueno (2010)]{manso10} but including macroscopic vertical motions. Lorentz damping rates, depolarizing elastic collisions, inelastic collisional rates and collisional alignment transfer rates were obtained from \cite[{Shine} \& {Linsky}(1974)]{Shine:1974aa}, \cite[Derouich et al (2007)]{Derouich:2007aa} and \cite[Manso Sainz \& Trujillo Bueno (2010)]{manso10}. During the iteration the SEE distribute the atomic level populations ($\propto \rho^0_0(\mathrm{J})$) among the corresponding magnetic energy sublevels. This creates the alignment terms $\rho^2_0(\mathrm{J})$, powered by the radiation field anisotropy, and the quantum coherences quantified by $\rho^2_Q(\mathrm{J})$ (with $Q\neq0$) and driven by the magnetic field kernel (Hanle effect). Finally, the RT equation for Stokes V Zeeman is solved independently without atomic polarization.
\section{Results: polarization maps and Hanle polarity inversion lines}
\label{sec:results}
\begin{figure}[t]
\begin{center}
\medskip
\smallskip
\includegraphics[scale=0.6]{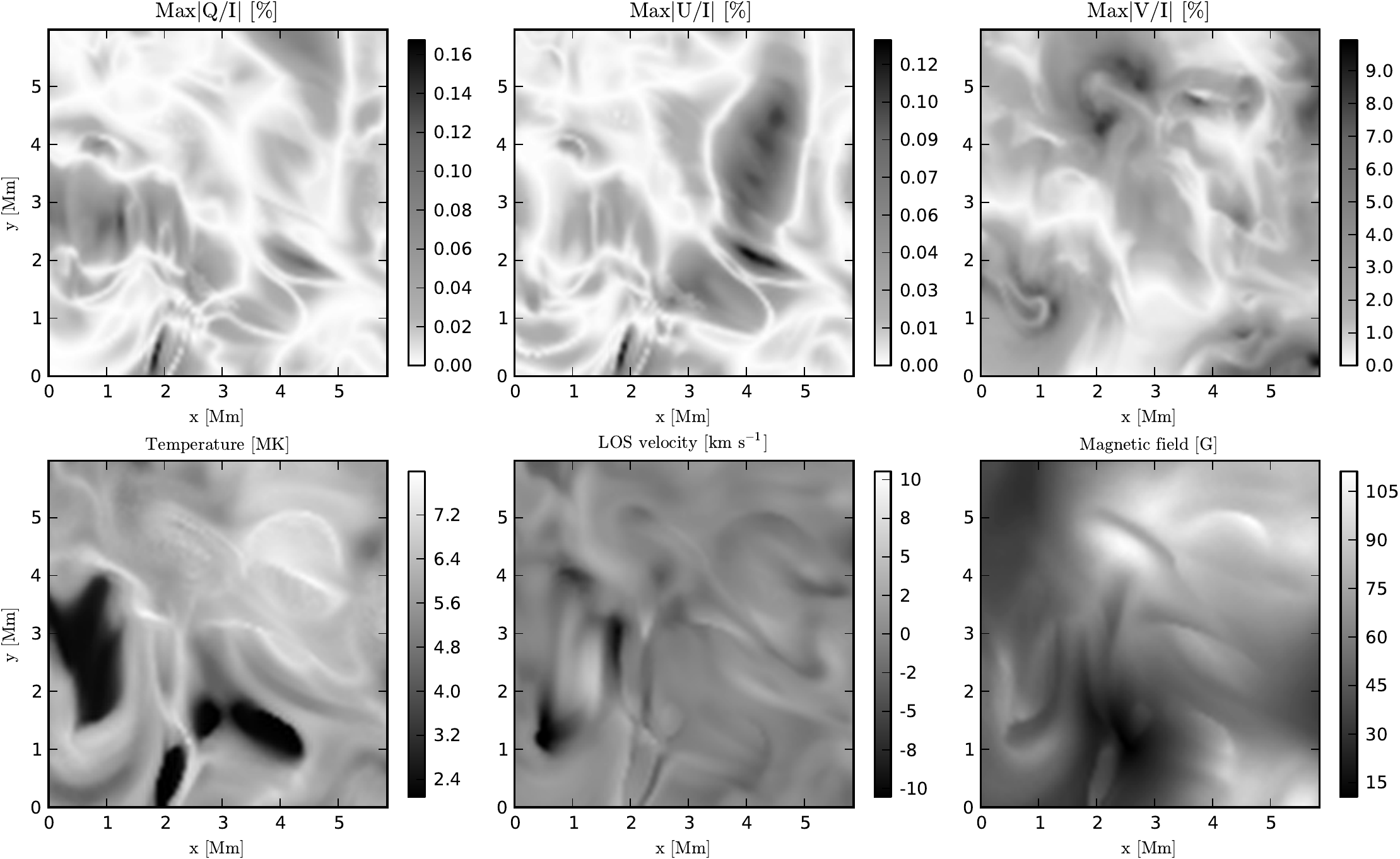}
\smallskip
\caption{Top: maximum fractional linear and circular polarization. Bottom: main physical quantities at heights with $\tau_{8542}=1$. The field of view is $\sim 8'' \times 8''$.}
\label{fig:fig1}
\end{center}
\end{figure}

Our results in the Ca {\sc ii} $8542$ {\AA} line illustrate the spatial behavior of the LP amplitudes in a quiet Sun context (see Fig. \ref{fig:fig1}). Note first that the LP in the maps is only significant in the presense of inclined magnetic fields which break the scattering symmetry (forward-scattering Hanle effect). The largest LP amplitudes are located in pixels having the largest vertical velocities in the horizontal field region. In general the amplitudes have the same order of magnitude as those calculated by \cite[Manso Sainz \& Trujillo Bueno (2010)]{manso10} in semi-empirical static models but, in regions where chromospheric velocities are above $\mathrm{\sim 5 \,km \cdot s^{-1}}$, the Doppler-enhanced polarization increases by one order of magnitude. Such maximum signals fill small patches in the maps and, because we analyzed a relatively small temporal snapshot, their real statistical significance is still not conclusive. On the other hand, these models are known to have a reduced kinematic in comparison with the real Sun \cite[(Leenaarts et al., 2009)]{Leenaarts:2009}, so the Doppler-induced enhancements of LP could be understimated. Pixels with maximum/negligible circular polarization are where the magnetic field \textit{below} the main formation heights of the line core is almost vertical/horizontal. Maximum Stokes V signals tend to coincide with minimum LP and viceversa. 
\subsection{Forward-scattering Hanle effect in saturation}\label{sec:fssat}
 The resulting LP is in the saturation regime of the Hanle effect because the chromospheric magnetism in these models is significantly stronger than the corresponding critical Hanle fields of the atomic levels involved in the $8542$ {\AA} transition. This is something that can easily happen in the real quiet Sun because solar weak fields are strong enough for saturation to occur in many lines. The consequence of Hanle saturation is that the LP becomes independent on the magnetic field strenght, but not on its orientation. The behavior of the line-center forward-scattering polarization in saturation can be explained with the following approximated expressions \footnote{These equations were deduced from Eqs. (7.16) of LL04 in the local reference frame of the solar atmosphere, whose $z$ axis points upwards along the solar radial and $x$ is parallel to the $x$ axis in our maps. Stokes Q is positive along $x$. In $\mathcal{F}$ we follow a standard notation for the upper and lower level quantum numbers  ($\alpha_u J_u$ and $\alpha_{\ell} J_{\ell}$) and the
  polarizability coefficients $\omega^{(K)}_{J_u J_{\ell}}$  \cite[({{Landi Degl'Innocenti} 1984})]{Landi-DeglInnocenti:1984}.}:
\begin{subequations}\label{eq:step6}
\begin{empheq}{align}
\frac{Q}{I} &\simeq \, -\frac{3}{4\sqrt{2}}\cdot\left[\sin^2{\theta_B}\cdot(3\cos^2{\theta_B}-1)\cos{(2\chi_B)}\cdot\mathcal{F}\right]_{\mathrm{\tau^{los}_{\nu_0}=1}}  \\
\frac{U}{I} &\simeq \,-\frac{3 }{4\sqrt{2}}\cdot\left[\sin^2{\theta_B}\cdot(3\cos^2{\theta_B}-1) \sin{(2\chi_B)}\cdot\mathcal{F}\right]_{\mathrm{\tau^{los}_{\nu_0}=1}},
 \end{empheq}
 \end{subequations}
where $\mathcal{F}=\omega^{(2)}_{J_u J_{\ell}} \sigma^2_0(J_u)
-\omega^{(2)}_{J_{\ell} J_u} \sigma^2_0(J_{\ell})$ is the non-magnetic
contribution of the fractional atomic alignment
($\sigma^2_0=\rho^2_0/\rho^0_0$) generated in the levels of the
transition. $\mathcal{F}$ is a sort of non-magnetic thermodynamical factor because it depends indirectly on kinematics and thermodynamics via the anisotropy. The angles $\chi_B$ and $\theta_B$ are the azimuth and inclination of the magnetic field vector in the solar atmospheric reference frame. Equations (\ref{eq:step6}) come from the $f_3$ term in Eq.(\ref{eq:one}): they explain what happens when the solar curvature effects are avoided. 
Note that the Hanle effect in forward scattering produces linearly polarized radiation being maximum along or perpendicularly to the projection of the magnetic field vector on the solar surface. Note also that Stokes Q and U are equivalent in their physical dependencies and have same maximum and minimum values, which does not occur in other LOS. Although being approximated expressions, they explain reasonably well the spatial patterns found in Fig. \ref{fig:fig1}. Hence, instead of being a drawback, the saturation of the Hanle effect could be of help for solar diagnosis.
\begin{figure}[t]
\begin{center}
\medskip
\smallskip
\includegraphics[scale=0.7]{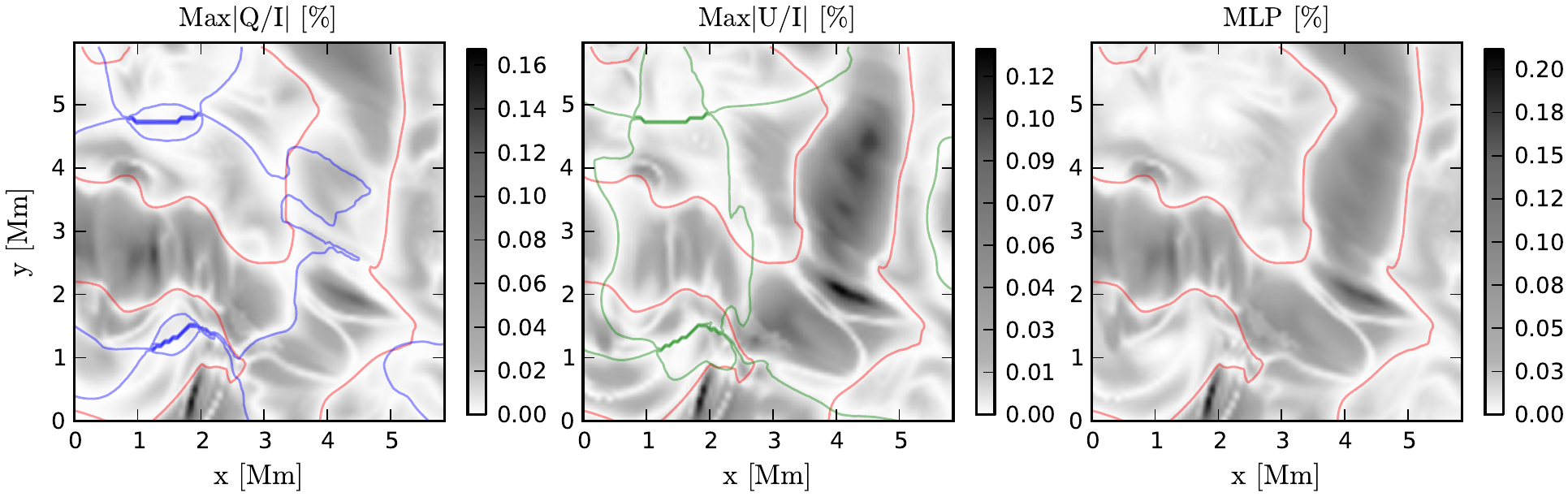}
\smallskip
\caption{Maximum fractional linear polarization in Q and U and MLP$= \sqrt{\mathrm{Max}^2|Q/I|+\mathrm{Max}^2|U/I|}$. Stokes Q is positive along $x$. Solid iso-contours trace constant $\chi_B$ (green and blue lines) and constant $\theta_B$ (red lines) in $\tau_{8542}=1$ for the field orientations expected in magnetic HPILs (see Sec. \ref{sec:hpils}). }
\label{fig:fig2}
\end{center}
\end{figure}
\subsection{Hanle polarity inversion lines}\label{sec:hpils}
A new idea emerging from the spatial topology of our synthetic maps is the concept of Hanle polarity inversion lines (HPILs): grooves where the fractional scattering polarization is zero. They are analogous to the polarity inversion lines in circular polarization. The HPILs appear in the maps due to the weak-field dependences in Q and U: they correspond to the zeroes of Eq. (\ref{eq:step6}). Consequently, the HPILs encode the magnetic field  topology across the formation region. 

We identify three kinds of HPILs, produced by three different sources, that can act together. 
The first kind of Hanle PIL connects pixels with the same magnetic field inclination and appear in the same place for Stokes Q and for U (hence also in the total linear polarization; see iso-contours in right panel of Fig. \ref{fig:fig2}). Where chromospheric magnetic fields are mostly-vertical ($\theta_B=  90 \pm 90^{\circ}$), these lines create wider spots without LP in the maps. They also appear separating vertical from horizontal field areas, namely where $\theta_B= 90 \pm 35.27^{\circ}$. As a magnetic field with $\theta_B=90 \pm 35.27^{\circ}$ forms the Van Vleck angle with the vertical, we call this first type as Van Vleck HPILs. Magnetic field emerging in small bipolar structures traces continuous Van Vleck HPILs enclosing the magnetic ``poles".

The HPILs can also be of \textit{azimuthal} kind if they are located where $\chi_B= 0^{\circ}, \pm 90^{\circ}, 180^{\circ}$ (for which $\mathrm{U}=0$; see iso-contours in mid panel of Fig.~\ref{fig:fig2}) or $\chi_B= \pm45^{\circ}, \pm135^{\circ}$ (for which $\mathrm{Q}=0$; see iso-contours in left panel of Fig.~\ref{fig:fig2}). Hence, pixels defining an azimuthal HPIL in Stokes Q (U) have a magnetic field vector lying either along or perpendicular to the reference direction for positive U (Q). Note that a Hanle PIL appearing in Stokes Q (U) is of azimuthal type if it does not appear in the same place for Stokes U (Q). Note also that, when azimuthal HPILs intersect, the cross point must have a magnetic field completely vertical (the cross ``point'' is a HPIL of the first kind). In other words, azimuthal HPILs have a radial nature, connecting areas of increased photospheric magnetic flux.

It is important to note that the iso-contours in Fig.~\ref{fig:fig2} are not approximated but calculated with the real chromospheric magnetic field orientation existing in the models at $\tau_{8542}=1$. Note then that the mere visual identification of a Van-Vleck/azimuthal HPIL is an accurate measurement of the field inclination/azimuth at the main formation height of a spectral line.

Finally, a third possible origin of HPILs are persisting spatial configurations of the radiation field anisotropy that nullifies the thermodynamical factor $\mathcal{F}$. Such \textit{thermodynamical} HPILs are co-spatial in Q and U, like the Van Vleck ones, but do not depend on magnetic field inclination. As illustrated in \cite[Carlin \& Asensio Ramos (2015)]{Carlin:2015a}, these lines seem to be separating regions having opposite kinematics (ascending vs. descending plasma) and thermodynamic (cool vs. hot plasma). This particular situation seems to happen around cool plasma volumes emerging through the chromosphere, either pushed by shock-driven convection or by flux emergence processes. Note how, once the Van-Vleck HPILs separating vertical from horizontal field in the total LP are identified (iso-contours in right panel of Fig.~\ref{fig:fig2}), the remaining HPILs are thermodynamical. The latter are abundant, sightly thinner and we speculate that, due to the emergence of shocks, perhaps less persistent in time than the magnetic ones. 
\section{Conclusions}
\label{sec:conclusions}
We have presented the landscape of a dynamic chromosphere where different processes creating spatial symmetry breakings (hence, also atomic and scattering polarization) compete for enhancing the linear polarization. The result of such contest is a natural cancellation of polarization in certain places, which creates a fingerprint made of intersecting null-polarization lines (HPILs, see also Carlin \& Asensio Ramos, 2015). When the solar curvature does not impose a preferential direction (this is, at the disk center), those features are dominated by magnetic field (Van-Vleck and azimuthal HPILs) and kinematics (thermodynamical HPILs), being also easier to match them with a physical driver. Thus, HPILs suggest a way of measuring/discriminating magnetic fields and kinematics without relaying in intensity proxies and, perhaps more importantly, only by visual inspection. Imaging spectropolarimetry then becomes indispensable. 

However, although the required spatial resolution can be achieved, the detection of the HPILs still demands good contrast and sensitivity, which is a problem for the $8542$ {\AA} line in current telescopes. We need to overcome the critical sensitivity gap existing between an effective line-core photon noise of $10^{-3}$ (where maximum disk-center signals can be attained) and $10^{-4}$ (where our results show clear HPILs in this spectral line). 

In summary, at disk center we lose photons polarized by scattering but if the spectral line and the instrumentation is adequate we could gain much in return.


\begin{thebibliography}{}

\bibitem[Anusha et~al. (2011)]{Anusha11}    
{Anusha, L.~S., Nagendra, K.~N., Bianda, M., Stenflo, J.~O., Holzreuter, R.,
  Sampoorna, M., Frisch, H., Ramelli, R., \& Smitha, H.~N.} 2011, 
\textit{ApJ} 737, 95

\bibitem[{{Bianda} {et~al.} (2011)}]{michele2011cai}
{Bianda}, M., {Ramelli}, R., {Anusha}, L.~S., {Stenflo}, J.~O., {Nagendra},
  K.~N., {Holzreuter}, R., {Sampoorna}, M., {Frisch}, H., \& {Smitha}, H.~N.
  2011, \textit{Astron. Lett.} 530, L13

\bibitem[Bommier\ (1997)]{bommier97}    
{Bommier, V.} 
1997, \textit{A\&A} 328, 726
\bibitem[{{Bommier} (2009)}]{Bommier:2009aa}   
{Bommier}, V. 2009, in: S.~V. {Berdyugina}, K.~N. {Nagendra}, \& R.~{Ramelli} 
(eds.), \textit{Solar Polarization 5: In Honor of Jan Stenflo}, 
ASP Conf. Series 405 (San Francisco: ASP), p. 335

\bibitem[{Carlin} {et~al.} (2013)]{Carlin:2013aa}   
{Carlin}, E.~S., {Asensio Ramos}, A., \& {Trujillo Bueno}, J. 2013, \textit{ApJ} 764, 40

\bibitem[{Carlin} \& Asensio Ramos (2015)]{Carlin:2015a}   
{Carlin}, E.~S.  \& {Asensio Ramos}, A. 2015, \textit{ApJ} 801, 16

\bibitem[{{Carlsson} \& {Stein}(1997)}]{carlsson_stein97}
{Carlsson}, M., \& {Stein}, R.~F. 1997, \textit{ApJ}, 481, 500

\bibitem[{{Derouich} {et~al.} (2007)}]{Derouich:2007aa}
{Derouich}, M., {Trujillo Bueno}, J., \& {Manso Sainz}, R. 2007, \textit{A\&A} 
472, 269


\bibitem[{{Hansteen} {et~al.} (2007)}]{Hansteen:2007}
{Hansteen}, V.~H., {Carlsson}, M., \& {Gudiksen}, B. 2007, in: 
P.~{Heinzel}, I.~{Dorotovi{\v c}}, \& R.~J. {Rutten} (eds.), 
\textit{The Physics of Chromospheric Plasmas}, 
ASP Conf. Series 368 (San Francisco: ASP), p. 107


\bibitem[{{Holzreuter} {et~al.} (2005)}]{Holzreuter:2005aa}
{Holzreuter}, R., {Fluri}, D.~M., \& {Stenflo}, J.~O. 2005, \textit{A\&A} 
434, 713
 
 \bibitem[{{Landi Degl'Innocenti} (1984)}]{Landi-DeglInnocenti:1984}
{Landi Degl'Innocenti}, E. 1984, \textit{Solar Phys.} 91, 1

\bibitem[Landi Degl'Innocenti \& Landolfi\ (2004)]{ll04}   
{Landi Degl'Innocenti, E., \& Landolfi, M.} 2004, \textit{Polarization in 
spectral lines}, Kluwer, Dordrecht (LL04)

\bibitem[{{Leenaarts} {et~al.} (2009)}]{Leenaarts:2009}
{Leenaarts}, J., {Carlsson}, M., {Hansteen}, V., \& {Rouppe van der Voort}, L.
  2009, \textit{Ap. Lett.} 694, L128

\bibitem[{{Manso Sainz} \& {Trujillo Bueno} (2003)}]{Manso-Sainz:2003}    
{Manso Sainz}, R., \& {Trujillo Bueno}, J. 2003, in: J.~{Trujillo-Bueno} \& 
J.~{Sanchez Almeida} (eds.), \textit{Solar Polarization}, ASP Conf. 
Series 307 (San Francisco: ASP), p. 251

\bibitem[Manso Sainz \& Trujillo Bueno\ (2010)]{ms10}   
{Manso Sainz, R., \& Trujillo Bueno, J.} 2010, \textit{ApJ} 722, 1416

\bibitem[{{Mili{\'c}} \& {Faurobert}(2012)}]{Milic:2012aa}   
{Mili{\'c}}, I., \& {Faurobert}, M. 2012, \textit{ApJ}, 539, A10

\bibitem[{Shine} \& {Linsky}(1974)]{Shine:1974aa}
{Shine}, R.~A., \& {Linsky}, J.~L. 1974, \textit{Solar Physics}, 39, 49

\bibitem[Stenflo \& Keller\ (1997)]{sk97}     
{Stenflo, J. O., \& Keller, C.~U.} 1997, \textit{A\&A} 321, 927

\bibitem[{{Stenflo}(2006)}]{Stenflo:2006aa}     
{Stenflo}, J.~O. 2006, in: R.~{Casini} \& B.~W. {Lites} (eds.), \textit{Solar 
Polarization 4}, ASP Conf. Series 358 (San Francisco: ASP), p. 215

\bibitem[Trujillo~Bueno (2001)]{jtb01}      
{Trujillo~Bueno, J.} 2001, in: M.~Sigwarth (ed.), \textit{Advanced Solar 
Polarimetry -- Theory, Observation, and Instrumentation}, ASP
ASP Conf. Series 236 (San Francisco: ASP), p.\ 161 

\bibitem[{{Trujillo Bueno} (2003)}]{Trujillo-Bueno:2003aa}              
{Trujillo Bueno}, J. 2003, in: J.~{Trujillo-Bueno} \& J.~{Sanchez Almeida} 
(eds.), \textit{Solar Polarization}, ASP Conf. Series 307 
(San Francisco: ASP), p.~407

\end{thebibliography}
\end{document}